\documentclass[prd,reprint,nofootinbib,superscriptaddress,preprintnumbers,showpacs]{revtex4-1}
\usepackage{amsfonts}
\usepackage{amssymb}
\usepackage{amsmath}
\usepackage{graphicx,color}
\usepackage{dcolumn}
\usepackage{epsfig}
\usepackage{bm}
\usepackage{bbm}
\usepackage{ulem}
\usepackage{hyperref}
\usepackage{lineno}
\usepackage{float}
\usepackage{subfigure}

\allowdisplaybreaks[4]

\pdfoutput=1

\begin{document}
	
	\title{Triangle mechanism in the decay process $B_0 \to J/\psi K^0 f_0(980)(a_0(980))$}
	\author{Jialiang Lu}
	\affiliation{School of Physics and Optoelectronics Engineering, Anhui University, Hefei 230601, People's Republic of China }
	\author{Xuan Luo}	
	\email{xuanluo@ahu.edu.cn}
    \affiliation{School of Physics and Optoelectronics Engineering, Anhui University, Hefei 230601, People's Republic of China }
	\author{Mao Song}
	\affiliation{School of Physics and Optoelectronics Engineering, Anhui University, Hefei 230601, People's Republic of China }
	\author{Gang Li}
	\affiliation{School of Physics and Optoelectronics Engineering, Anhui University, Hefei 230601, People's Republic of China }	
	\begin{abstract} 
		\vspace{0.5cm}
	The role of the triangle mechanism in the decay process $B_0\to J/\psi K^0f_0 \to J/\psi K^0\pi^+\pi^-$ and $B_0\to J\psi K^0a_0\to J/\psi K^0 \pi^0\eta$ is probed. In these process, the triangle singularity appears from the decay of $B^0$ into $J/\psi\phi K^0$ then $\phi$ decays into $K^0\bar{K^0}$ and $K^0\bar{K^0}$ merged into $f_0$ or $a_0$ which finally decay into $\pi^+\pi^-$ and $\pi^0\eta$ respectively. We find that this mechanism leads to a triangle singularity around $M_{\rm inv}(K^0f_0(a_0))\approx1520\ {\rm MeV}$, and gives sizable branching fractions ${\rm{Br}}(B_0\to J/\psi K^0f_0\to J/\psi K^0\pi^+\pi^-)=1.38\times10^{-6}$ and $ {\rm{Br}}(B_0\to J/\psi K^0a_0\to J/\psi K^0\pi^0\eta)=2.56\times10^{-7}$. This potential investigation can help us obtain the information of the scalar meson $f_0(980)$ or $a_0(980)$.	
	\end{abstract}
	\maketitle	
	\section{Introduction}
	\label{I}
	
The dynamics of the strong interaction are described by quantum chromodynamics (QCD), and hadron spectroscopy is a method of studying QCD, meanwhile, hadron spectroscopy is the basic theory of the strong interaction. All the time, understanding the spectrum of hadron resonances \cite{ParticleDataGroup:2016lqr} and establishing a connection with the QCD, is one of the important goals of hadron physics. The conventional quark models in the low-lying hadron spectrum successfully explain that baryon is a complex of three quarks, and meson is a combination of quark and antiquark \cite{Godfrey:1985xj,Capstick:1986ter}. However, even if the model provides a great deal of data about the meson and baryon resonances \cite{Godfrey:1985xj,Capstick:1986ter,Vijande:2004he}, we also cannot rule out other more exotic components, especially considering that the QCD Lagrangian includes not only quarks but gluons as well. This leads to other configurations of color singlets, such as glueballs made purely of gluons, mixed states made of a quark and gluon excitations, and multiquarks. There is also the possibility of more quark states in a hadron, such as $qq\bar{q}\bar{q}$ and $qq\bar{q}qq$, which were already mentioned in the Ref.\  \cite{Gell-Mann:1964ewy}. To understand quantitatively the QCD of quarks and gluons, over the years, numerous related experiments have been conducted in search of evidence for these exotic components in the mesonic and baryonic spectrum \cite{Klempt:2007cp,Crede:2008vw,Brambilla:2014jmp,Chen:2016qju,Guo:2017jvc}. 

The triangle singularity (TS) has been discussed in Ref.\  \cite{Karplus:1958zz}, and Landau has systematized it in Ref.\  \cite{Landau:1959fi}. The TS was fashionable in the 1960s \cite{Peierls:1961zz,Aitchison:1964zz,Bronzan:1964zz,Schmid:1967ojm}. In addition to ordinary hadronic, molecular, or multiquarks states, TSs can produce peaks, but these peaks are produced by simple kinematic effects. The Coleman-Norton theorem \cite{Coleman:1965xm} states that the Feynman amplitude has a singularity on the physical boundary as time moves forward if the decay process can be interpreted as taking place during the conservation of energy and momentum in space-time, and all internal particles really exist on the shell. In the process of particle 1 decays into particles 2 and 3, particle 1 first decays into particles A and B, then A decays into particles 2 and C, and finally particles B and C fused  into an external particle 3. Particles A, B, and C are intermediate particles, and if the momenta of these intermediate particles can take on-shell values, a singularity will occur. A new way to understand this process is proposed in Ref.\ \cite{Bayar:2016ftu}, this method does not compute the entire amplitude of the Feynman diagram including the triangle loop. The condition for producing a TS is $q_{on+}=q_{a-}$ \cite{Bayar:2016ftu}, where $q_{on}$ is the on-shell momentum of particle A or B in the rest system of particle 1, $q_{a-}$ defines one of the two solutions to the momentum of particle B when B and C produce particle 3 on the shell. Since the process of the triangle mechanism involves the fusion of hadrons, the presence of hadronic molecular states plays an important role in having measurable strength. Therefore, the study of singularity is also a useful tool to study the molecular states of hadrons.

The isospin violation in production of the $f_0(980)$ or $a_0(980)$ resonances generation and its mixing have long been controversial. We have to give up the idea of trying to establish a "$f_0-a_0$ mixing parameter" out of different reactions because it was shown that the isospin violation depends a lot on the reaction \cite{Wu:2007jh,Hanhart:2007bd,Wu:2011yx,Aceti:2012dj}. The spark was raised by the puzzle of the anomalously large isospin violation in the $\eta(1450)\to\pi^0f_0(980)$ decay \cite{BESIII:2012aa}, which was due to a TS and explained in Refs.\ \cite{Wu:2011yx,Aceti:2012dj}. From the point of view of the $f_0(980)$ and $a_0(980)$ themselves, the above reaction is also very enlightening. These resonances are not produced directly but are produced by the rescattering of $K\bar{K}$, the mechanism for the formation of these resonances in the chiral unitary approach \cite{Oller:1997ti,Kaiser:1998fi,Locher:1997gr,Nieves:1999bx}. In addition, there are a large number of processes with the same mechanism, such as the process $\tau^-\to\nu_{\tau}\pi^-f_0 (a_0)$ \cite{Dai:2018rra}, the process $B^0_s\to J/\psi\pi^0f_0(a_0)$ \cite{Liang:2017ijf}, the process $D_s^+\to\pi^+\pi^0f_0(a_0)$ \cite{Sakai:2017iqs}, the process $B^-\to D^{\ast 0}\pi^-f_0(a_0)$ \cite{Pavao:2017kcr} and the same TS was shown in Refs.\ \cite{Mikhasenko:2015oxp,Aceti:2016yeb} to provide a plausible explanation for the peak observed in the $\pi f_0(980)$ final state, it is easy to envisage many reactions of this type \cite{Liang:2019jtr}, it also inspires us to find more processes like this type and search for TS enhanced isospin-violating reactions producing the $f_0(980)$ or $a_0(980)$ resonances \cite{Liang:2019yir}. 

In the present work we study the reactions $B^0\to J/\psi K^0f_0(980)$ and $B^0\to J/\psi K^0a_0(980)$,  both decays modes are allowed. The process followed by the $\phi$ decay into $K^0\bar{K}^0$ and the $K^0\bar{K}^0$ fuse into $f_0(a_0)$ generate a singularity, we show that it develops a TS at an invariant mass $M_{\rm inv}(K^0R)\simeq1520\ {\rm MeV}$. Meanwhile, we can obtain $\frac{d^2\Gamma}{dM_{\rm inv}(K^0f_0)dM_{\rm inv}(\pi^+\pi^-)}$ or $\frac{d^2\Gamma}{dM_{\rm inv}(K^0a_0)dM_{\rm inv}(\pi^0\eta)}$ which show the shapes of the $f_0$(980) and $a_0$(980) resonances in the $\pi^+\pi^-$ or $\pi^0\eta$ mass distributions respectively. We can restrict the integral in $M_{\rm inv}(\pi^+\pi^-)$ or $M_{\rm inv}(\pi^0\eta)$ to this region when calculating the mass distribution $\frac{d^2\Gamma}{dM_{\rm inv}(K^0f_0)dM_{\rm inv}(\pi^+\pi^-)}$ or $\frac{d^2\Gamma}{dM_{\rm inv}(K^0a_0)dM_{\rm inv}(\pi^0\eta)}$. Then we integrate over the $\pi^+\pi^-$ or $\pi^0\eta$ invariant masses and obtain which shows a clear peak around $M_{\rm inv}(K^0R)\simeq1520\ {\rm MeV}$. The further integration over $M_{\rm inv}(K^0R)$ provides us branching ratios for $B^0\to J/\psi K^0\pi^+\pi^-$ $B^0\to J/\psi K^0\pi^0\eta$, we find that the mass distribution of these decay processes showed a peak associated with TS. In addition, the corresponding decay branching ratio is obtained, and we find the branching fractions ${\rm Br}(B_0\to J/\psi K^0f_0(a_0))=1.007\times10^{-5}$, ${\rm Br}(B_0\to J/\psi K^0f_0(980)\to J/\psi K^0\pi^+\pi^-)=1.38\times10^{-6}$ and ${\rm Br}(B_0\to J/\psi K^0a_0(980)\to J/\psi K^0\pi^0\eta)=2.56\times10^{-7}$.In any case, the main aim of the present work is to point out the presence of the TS in this reaction. This work provides one more measurable example of a TS, which has been quite sparse up to now. The singularity generated by this process can also play an early warning role for future experiments. 

	\section{Framework}
	\label{II}
	\begin{figure}
		\centering 
		\includegraphics[width=0.4\textwidth]{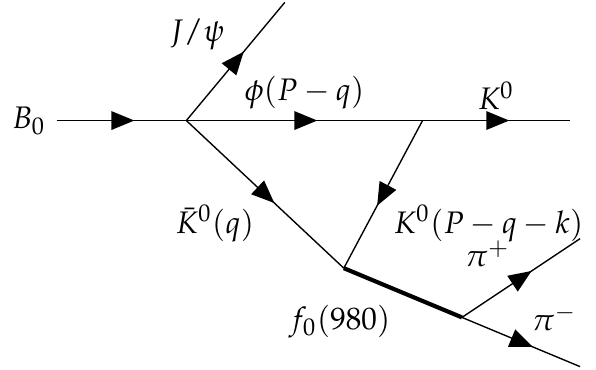}
		\caption{The Feynman diagrams of the decay process $B_0 \to J/\psi K^0 f_0(980)\to J/\psi K^0\pi^+ \pi^-$ involving a triangle loop.} 
		\label{fig1} 
	\end{figure}
	\begin{figure}
	\centering 
	\includegraphics[width=0.4\textwidth]{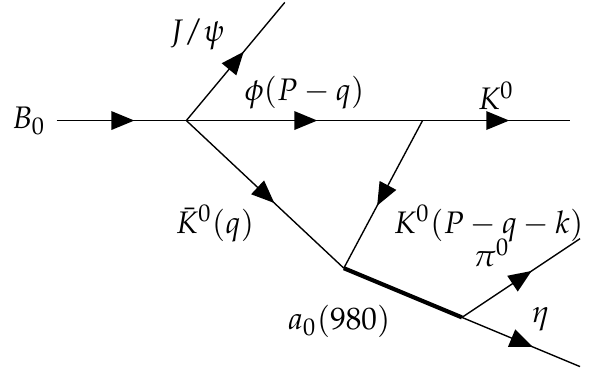}
	\caption{The Feynman diagrams of the decay process $B_0 \to J/\psi K^0 a_0(980)\to J/\psi K^0\pi^0 \eta$ involving a triangle loop.} 
	\label{fig2} 
	\end{figure}

We plot the Feynman digrams of the decay process $B_0 \to J/\psi K^0 f_0(a_0)(980)$ involving a triangle loop in Fig.\,\ref{fig1} and Fig.\,\ref{fig2}, we can observe that particle $B_0$ first decays into particles $J/\psi, \phi, K^0$ and then particle $\phi$ decays into $K^0$ and $\bar{K}^0$. It is worth noting that $K^0$ is faster than $\bar{K}^0$, this causes $K^0$ and $\bar{K}^0$ fuse to $f_0(a_0)$, and eventually  $f_0(a_0)$ decays into $\pi^+\pi^-(\pi^0\eta)$.

We take Fig.\,\ref{fig1} for example to perform the following discussion since Fig.\,\ref{fig1} and Fig.\,\ref{fig2} have nearly the same amplitude. Now we want to find the position of TS in complex-$q$ plane, instead of evaluating the whole amplitude of a Feynman diagram including triangle loop, analogously to Refs.\ \cite{Bayar:2016ftu,Wang:2016dtb,Huang:2020kxf} we can use
\begin{align}
	\label{1}
	q_{on+}=q_{a-}	,
\end{align}
where the $q_{on+}$   is the on-shell three momenta of the $K^0$ in the center of mass frame of $\phi K^0$, meanwhile, $q_{a-}$ defines one of the two solutions for the momentum of $\phi$ when $\phi \bar{K}^0$ are on-shell to produce $f_0$ and the $q_{a-}$ can be obtained by analyzing the singularity structure of the triangle loop. $q_{on+}$ and $q_{a-}$ are given by
\begin{align}
	q_{on+}=\frac{\lambda^{\frac{1}{2}}(s,M_{\phi}^2,M^2_{K^0})}{2\sqrt{s}},\qquad q_{a-}=\gamma(vE_{K^0}-p^*_{K^0})-i\epsilon,
\end{align}
we use $s$ denotes the squared invariant mass of $\phi$ and $K^0$, and $\lambda(x,y,z)=x^2+y^2+z^2-2xy-2yz-2xz$ is the k{\"a}hlen function.
with definition
\begin{align}
	v=\frac{k}{E_{f_0}},\qquad    \gamma=\frac{1}{\sqrt{1-v^2}}=\frac{E_{f_0}}{m_{f_0}},\notag\\
	E_{K^0}=\frac{m_{K^+_1}^2+m_{K^0}^2-m_{\bar{K}^0}^2}{2m_{f_0}},p^*_{K^0}=\frac{\lambda^{\frac{1}{2}}(m^2_{f_0},m_{K^0}^2,m_{\bar{K}^0}^2)}{2m_{f_0}}.
\end{align}

It is easy to realize that $E_{K^0}$ and $p^*_{K^0}$ are the energy and momentum of $K^0$ in the center of mass frame of the  $\phi K^0$ system, $v$ and $\gamma$ are the velocity of the $f_0$ and Lorentz boost factor. In addition, we can easily get
\begin{align}
	E_{f_0}=\frac{s+m^2_{f_0}-m^2_{\bar{K}^0}}{2\sqrt{s}},\ k=\frac{\lambda^{\frac{1}{2}}(s,m_{f_0}^2,m^2_{\bar{K}^0})}{2\sqrt{s}}.
\end{align}

The three intermediate particles must be on the shell, in the $\phi K^0$  center of mass frame, we let the mass of $f_0$ slightly larger than the mass sum of $K^0$ and $\bar{K}^0$, now we can determine that the mass of $K^0$ is $497.61\ \rm M\rm eV$, because we're making the mass of $f_0$ a little bit greater than the sum of the masses of $K^0$ and $\bar{K}^0$, so now we're making the mass of $f_0$ $991\ \rm{MeV}$, we can find a TS at around $\sqrt{s}=1520\ \rm{MeV}$. If we use in Eq.\,\ref{1} complex masses $(M-i\Gamma/2)$ of mesons which include widths of $K^0$ and $\bar{K}^0$, the solution of Eq.\,\ref{1} is then $(1520-10i)\ \rm{MeV}$ and the solution implies that the TS has a "width" $20\ \rm{MeV}$.

\subsection{The decay process $B_0\to J/\psi\phi K^0$}
Here, we only focus on the $B_0\to J/\psi\phi K^0$ process and the decay branching ratio has been experimentally measured to be Ref.\ \cite{ParticleDataGroup:2020ssz}
\begin{align}
	{\rm{Br}}(B_0\to J/\psi\phi K^0)=(4.9\pm 1.0)\times10^{-5}.
\end{align}
The differential decay width over the invariant mass distribution $\phi K^0$ can be written as
\begin{align}
	\frac{d\Gamma_{B_0\to J/\psi\phi K^0}}{d{M_{\rm{inv}}}(\phi K^0)}=\frac{1}{(2\pi)^3}\frac{|{\vec p^*_{K^0}}|| {\vec p_{J/\psi}}|}{4M_{B_0}^2}\cdot \overline{\sum}|t_{B_0\to J/\psi \phi K^0}|^2,
\end{align}
where 
\begin{align}
	|{\vec p^*_{K^0}}|&=\frac{\lambda^{\frac{1}{2}}(M_{\rm inv}(\phi K^0)^2,m_{\phi}^2,m_{K^0}^2)}{2M_{\rm inv}(\phi K^0)},\\\notag
	|{\vec p_{J/\psi}}|&=\frac{\lambda^{\frac{1}{2}}(M_{B_0}^2,m_{J/\psi}^2,M_{\rm inv}(\phi K^0)^2)}{2M_{B_0}}.\\
\end{align}
we use the polarization summation formula 
\begin{align}
	\sum\limits_{pol}=\epsilon_{\mu}(p)\epsilon_{\nu}(p)=-g_{\mu\nu}+\frac{p_{\mu}p_{\nu}}{m^2},
\end{align}
we have
\begin{align}
	\overline{\sum_{pol}}|t_{B_0\to J/\psi\phi K^0}|^2&=\mathcal{C}^2 \left( -g_{\mu\nu}+\frac{p_{\phi}^\mu p_{\phi}^\nu}{m_{\phi}^2} \right)\left( -g_{\mu\nu}+\frac{p_{K^0}^\mu p_{K^0}^\nu}{m_{K^0}^2} \right)
	\\\notag
	&=\mathcal{C}^2\left( 2+\frac{(p_{\phi} \cdot p_K)^2}{m_{\phi}^2 m_K^2} \right).
\end{align}
where 
\begin{align}
	p_{\phi}\cdot p_{K^0}=\frac{1}{2}(M_{\rm inv}(\phi K^0)^2-m_{\phi}^2-m_{K^0}^2),
\end{align}
\begin{align}
	\frac{1}{\Gamma_{B_0}}=\frac{{\rm{Br}}(B_0\to J/\psi\phi K^0)}{\int\frac{d\Gamma_{B_0}}{dM_{\rm inv}(\phi K^0)}dM_{\rm inv}(\phi K^0)},
\end{align}
Then one can obtain
\begin{align}
	\frac{\mathcal{C}^2}{\Gamma_{B_0}}=\frac{{\rm{Br}}(B_0\to J/\psi\phi K^0)}{\int dM_{\rm inv}(\phi K^0)\frac{1}{(2\pi)^3}\frac{|{\vec p^*_{K^0}}|| {\vec p_{J/\psi}}|}{4M_{B_0}^2}\cdot\left( 2+\frac{(p_{\phi} \cdot p_K)^2}{m_{\phi}^2 m_K^2} \right)},
\end{align}
where the integration is performed from $M_{\rm inv}(\phi K^0)_{\rm{min}}$=$m_{\phi}+m_{K^0}$ to $M_{\rm inv}(\phi K^0)_{\rm{max}}$=$M_{B_0}-m_{J/\psi}$.\\

\subsection{The triangle mechanism in the decay $B_0\to J/\psi K^0 f_0(980)\ ,\ f_0(980)\to \pi^+\pi^-$}
In the previous subsection, we have calculated the transition strength of the decay process $B_0\to J/\psi K^0 f_0(980)$, now we calculate the contribution of the vertex $\phi\to K^0\bar{K}^0$, we can obtain this $VPP$ vertex from the chiral invariant lagrangian with local hidden symmetry given in Refs.\ \cite{Bando:1984ej,Bando:1987br,Meissner:1987ge,Nagahiro:2008cv}. 
\begin{align}
	\label{vpp}
	\mathcal{L}_{VPP}=-ig\left\langle V^{\mu}[P,\partial_{\mu}P]\right\rangle,
\end{align}
where the $\left\langle\right\rangle$ stand for the trace of the flavor SU(3) matrices and $g$ is the coupling in the local hidden gauge ,
\begin{align}
	g=\frac{M_V}{2f_{\pi}},\qquad M_V=800\ {M\rm eV},\qquad {f_{\pi}}=93\ \rm M\rm eV.
\end{align}
The $V$ and $P$ in Eq.\,\ref{vpp} are the vector meson matrix and pseudoscalar meson matrix in the SU(3) group,  $P$ and $V$ are given by
\begin{align}
	\label{2}
	&P=\left(\begin{array}{ccc}
		\frac{\pi^{0}}{\sqrt{2}}+\frac{\eta}{\sqrt{3}}+\frac{\eta^{\prime}}{\sqrt{6}} & \pi^{+} & K^{+} \\
		\pi^{-} & -\frac{\pi^{0}}{\sqrt{2}}+\frac{\eta}{\sqrt{3}}+\frac{\eta^{\prime}}{\sqrt{6}} & K^{0} \\
		K^{-} & \bar{K}^{0} & -\frac{\eta}{\sqrt{3}}+\sqrt{\frac{2}{3}} \eta^{\prime}
	\end{array}\right),\notag\\
	&V=\left(\begin{array}{ccc}
		\frac{\rho^{0}}{\sqrt{2}}+\frac{\omega}{\sqrt{2}} & \rho^{+} & K^{*+} \\
		\rho^{-} & -\frac{\rho^{0}}{\sqrt{2}}+\frac{\omega}{\sqrt{2}} & K^{* 0} \\
		K^{*-} & \bar{K}^{* 0} & \phi
	\end{array}\right).
\end{align}
we get
	
	\begin{align}
		\label{tT}
		t_T&=\int \frac{d^3 q}{(2\pi)^3}\frac{1}{8\omega_{K^0}\omega_\phi \omega_{K^0}^{\prime}}\frac{1}{k^0-\omega_{K^0}^{\prime}-\omega_{\phi}}
		\notag\\
		&\times \frac{1}{M_{\rm inv}(K^0f_0)+\omega_{K^0}+\omega_{K^0}^{\prime}-k^0}
		\notag\\
		&\times \frac{1}{M_{\rm inv}(K^0f_0)-\omega_{K^0}-\omega_{K^0}^{\prime}-k^0+i\frac{\Gamma_{K^0}}{2}}
		\notag\\
		&\times \bigg[ \frac{2M_{\rm inv}(K^0f_0)\omega_{K^0}+2k^0\omega_{K^0}^{\prime}}{M_{\rm inv}(K^0f_0)-\omega_{\phi}-\omega_{K^0}+i\frac{\Gamma_{\phi}}{2}+i\frac{\Gamma_{K^0}}{2}} 
		\notag\\
		&-\frac{2(\omega_{K^0}+\omega_{K^0}^{\prime})(\omega_{K^0}+\omega_{K^0}^{\prime}+\omega_{\phi})}{M_{\rm inv}(K^0f_0)-\omega_{\phi}-\omega_{K^0}+i\frac{\Gamma_{\phi}}{2}+i\frac{\Gamma_{K^0}}{2}}\bigg]\left( 2+\frac{\vec q \cdot \vec k}{{\vec k}^2} \right),
	\end{align}
	where
	\begin{align}
		&\omega_{K^0}^{\prime}=((\vec{P}-\vec{q}-\vec{k})^2+m_{K^0}^2)^{\frac{1}{2}},\notag\\
		&\omega_{\phi}=((\vec{P}-\vec{q})^2+m_{\phi}^2)^{\frac{1}{2}},\notag\\
		&\omega_{K^0}=(\vec{q}^2+m_{K^0}^2)^{\frac{1}{2}},\notag\\
		&k^0=\frac{M_{\rm inv}^2(K^0f_0)+m_{K^0}^2-m_{f_0}^2}{2M_{\rm inv}(K^0f_0)},\notag\\
		&\left |\vec{k}\right |=\frac{\lambda^{\frac{1}{2}}(M^2_{\rm inv}(K^0f_0),m_{K^0}^2,m_{f_0}^2)}{2M_{\rm inv}(K^0f_0)}.
	\end{align}

Now we can write the invariant mass distribution $M_{\rm inv}(K^0f_0)$ in the decay $B_0\to J/\psi K^0 f_0$ as  
\begin{align}
	\frac{d\Gamma_{B_0\to J/\psi K^0 f_0}}{dM_{\rm inv}(K^0 f_0)}=\frac{1}{(2\pi)^3}\frac{1}{4M_{B^0}^2}p^{\prime}_{J/\psi}\tilde{p}^{\prime}_{K^0}\overline{\sum}\sum|t|^2,
\end{align}
where $p^{\prime}_{J/\psi}$ is the momentum of the $J/\psi$ in the $B_0$ rest frame , and $\tilde{p}^{\prime}_{K^0}=|\vec{k}|$ is the momentum of the $K^0$ in the $K^0f_0(980)$ rest frame , where 
\begin{align}
	p^{\prime}_{J/\psi}&=\frac{\lambda^{\frac{1}{2}}(M_{B_0}^2,m_{J/\psi}^2,M_{\rm inv}^2(K^0f_0))}{2M_{B_0}},\\	
	\tilde{p}^{\prime}_{K^0}&=\frac{\lambda^{\frac{1}{2}}(M_{\rm inv}^2(K^0f_0),m_{K^0}^2,m_{f_0}^2)}{2M_{\rm inv}(K^0f_0)}.
\end{align}
Then we have
\begin{align}
	\label{1to3}
	&\qquad\frac{1}{\Gamma_{B_0}}\frac{d\Gamma_{B_0\to J/\psi K^0 f_0}}{dM_{\rm inv}(K^0f_0)}\notag\\
	&=\frac{1}{(2\pi)^3}\frac{1}{4M_{B^0}}p^{\prime}_{J/\psi}\tilde{p}^{\prime}_{K^0}\frac{\mathcal{C}^2}{\Gamma_{B_0}}\cdot g^2g^2_{K^0\bar{K}^0f_0}|\vec{k}|^2|t_T|^2\notag\\
	&=\frac{1}{(2\pi)^3}\frac{1}{4M_{B^0}}p^{\prime}_{J/\psi}\tilde{p}^{\prime 3}_{K^0}\frac{\mathcal{C}^2}{\Gamma_{B_0}}\cdot g^2g^2_{K^0\bar{K}^0f_0}|t_T|^2.\\\notag
\end{align}

The $K^0\bar{K}^0\to \pi^+\pi^-$ and $K^0\bar{K}^0\to\pi^0\eta$ scattering has been studied in detail in Refs.\ \cite{Xie:2014tma,Liang:2014tia} within the chiral unitary approach, where altogether six channels were taken into account, including $\pi^+\pi^-$, $\pi^0\pi^0$ , $K^+K^-$ , $K^0\bar{K}^0$ , $\eta\eta$ and $\pi^0\eta$. In the present study we use this as input, and
we shall see simultaneously both the $f_0(980)$ (with $I$ = 0)
and $a_0(980)$ (with $I$ = 1) productions. 
Now we can write down the double differential mass distribution for the decay process $B_0\to J/\psi K^0f_0(980)\to J/\psi K^0\pi^+\pi^-$.
 
Now for the case of $f_0(980)$, we only have the decay $f_0\to\pi^+\pi^-$, and thus
\begin{align}
	\label{1to4}
	&\frac{d^2\Gamma_{B_0\to J/\psi K^0f_0(980)\to J/\psi K^0\pi^+\pi^-}}{dM_{\rm inv}(K^0f_0)dM_{\rm inv}(\pi^+\pi^-)}=\notag\\
	&\frac{1}{(2\pi)^5}\frac{1}{4M^2_{B_0}}p^{\prime\prime}_{J/\psi}\tilde p^{\prime\prime}_{K_0}\tilde p^{\prime\prime}_{\pi^+}\sum\overline{\sum}|t^{\prime}|^2,
\end{align}
where $p^{\prime\prime}_{J/\psi}$ is the momentum of the $J/\psi$ in the $B_0$ rest frame, $\tilde p^{\prime\prime}_{K^0}$ is the momentum of the $K^0$ in the $K^0f_0(980)$ ret frame, and $\tilde p^{\prime\prime}_{\pi}$ is the momentum of the $\pi$ in the $\pi^+\pi^-$ rest frame:
\begin{align*}
	&p^{\prime\prime}_{J/\psi}=\frac{\lambda^{\frac{1}{2}}(M_{B_0}^2,m_{J/\psi}^2,M_{\rm inv}^2(K^0f_0))}{2M_{B_0}},\notag\\
	&\tilde p^{\prime\prime}_{K_0}=\frac{\lambda^{\frac{1}{2}}(M_{\rm inv}^2(K^0f_0),m_{K^0}^2,M_{\rm inv}^2(\pi^+\pi^-))}{2M_{\rm inv}(K^0f_0)},\notag\\
	&\tilde p^{\prime\prime}_{\pi}=\frac{\lambda^{\frac{1}{2}}(M_{\rm inv}^2(\pi^+\pi^-),m_{\pi^+}^2,m_{\pi^-}^2)}{2M_{\rm inv}(\pi^+\pi^-)}.
\end{align*}
	\section{Results}
	\label{III}
\begin{figure*}
	\centering 
	\subfigure[]{\includegraphics[width=0.4\textwidth]{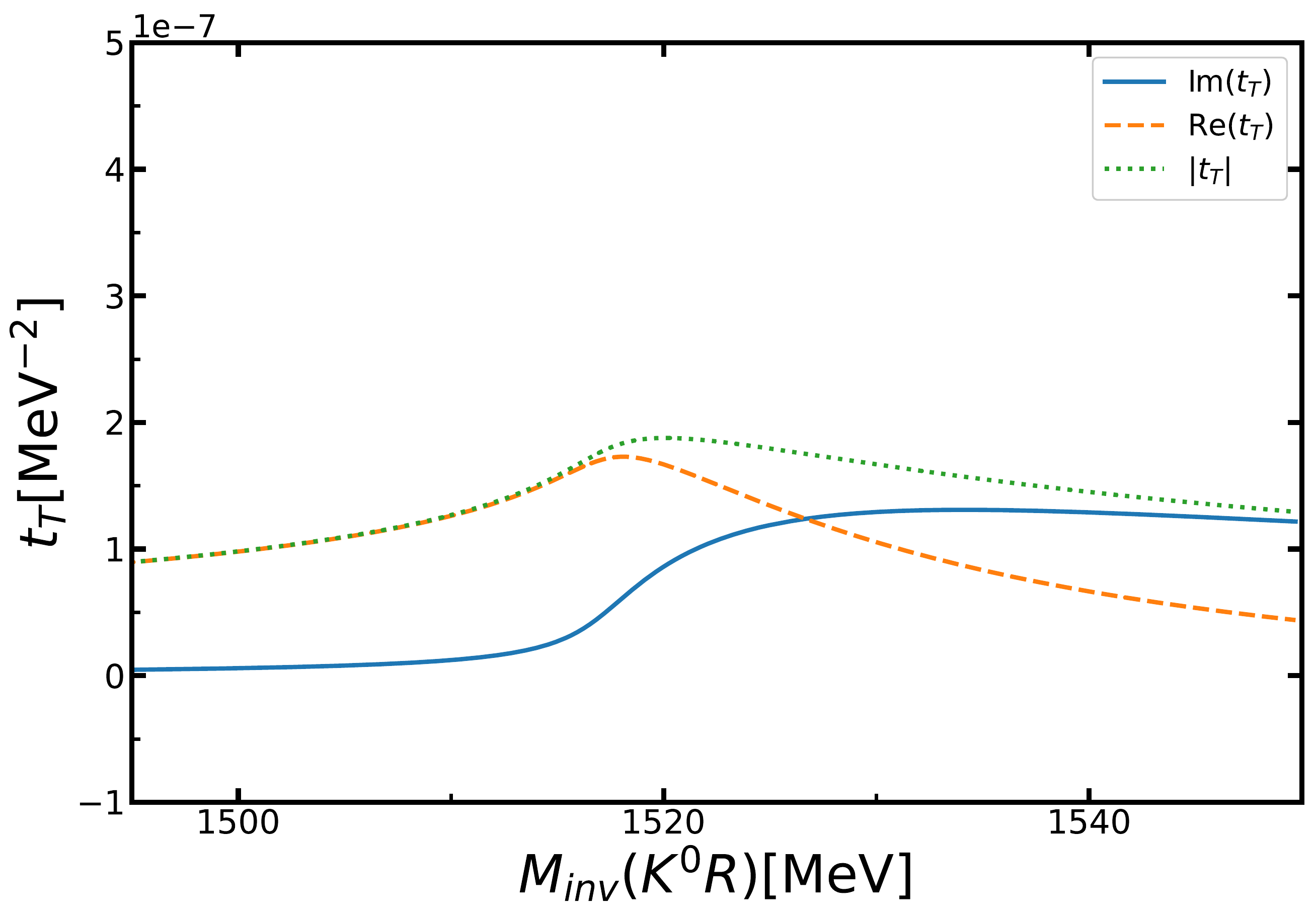}}
	\subfigure[]{\includegraphics[width=0.4\textwidth]{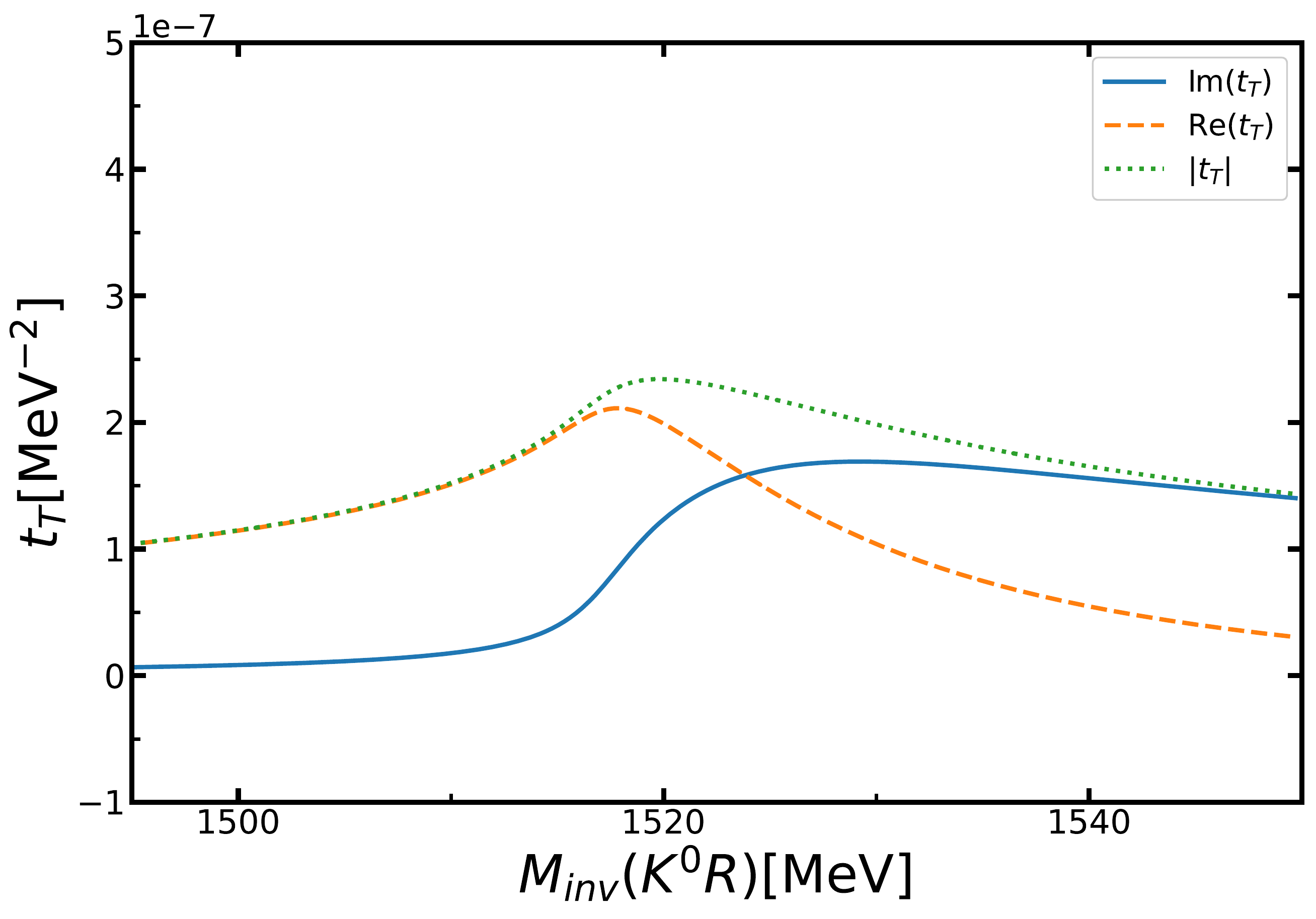}}
	\subfigure[]{\includegraphics[width=0.4\textwidth]{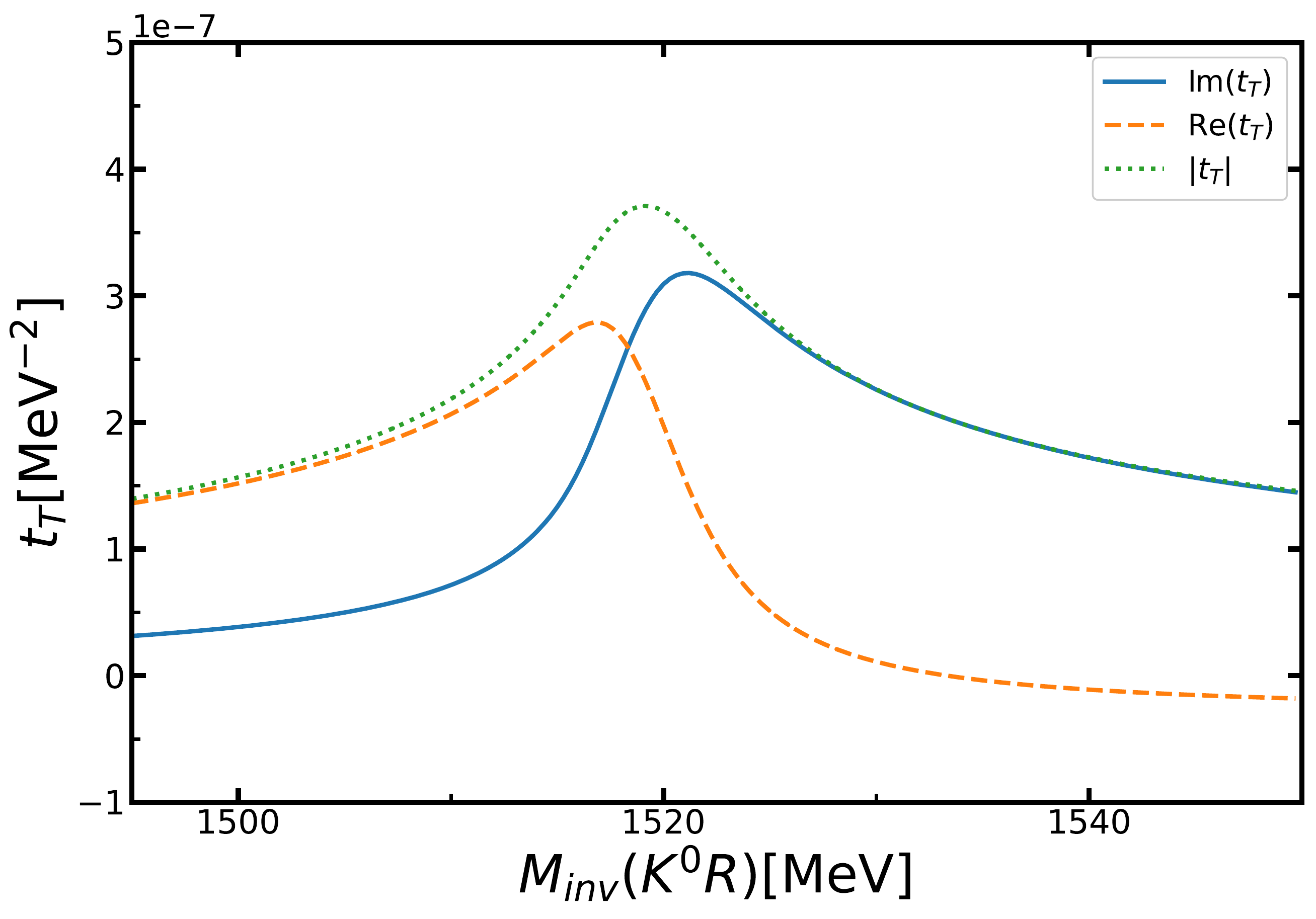}} 
	\caption{Triangle amplitude $t_T$, as a function of $M_{\rm inv}(K^0f_0/K^0a_0)$ for (a) $m_{f_0(a_0)}$=986 MeV, (b) $m_{f_0(a_0)}$=991 MeV and (c) $m_{f_0(a_0)}$=996 MeV. $|t_T|$, Re($t_T$) , and Im($t_T$) are plotted using the green, orange, and blue curves, respectively.} 
	\label{fig3} 
\end{figure*}
\begin{figure*}
	\centering 
	\includegraphics[width=0.4\textwidth]{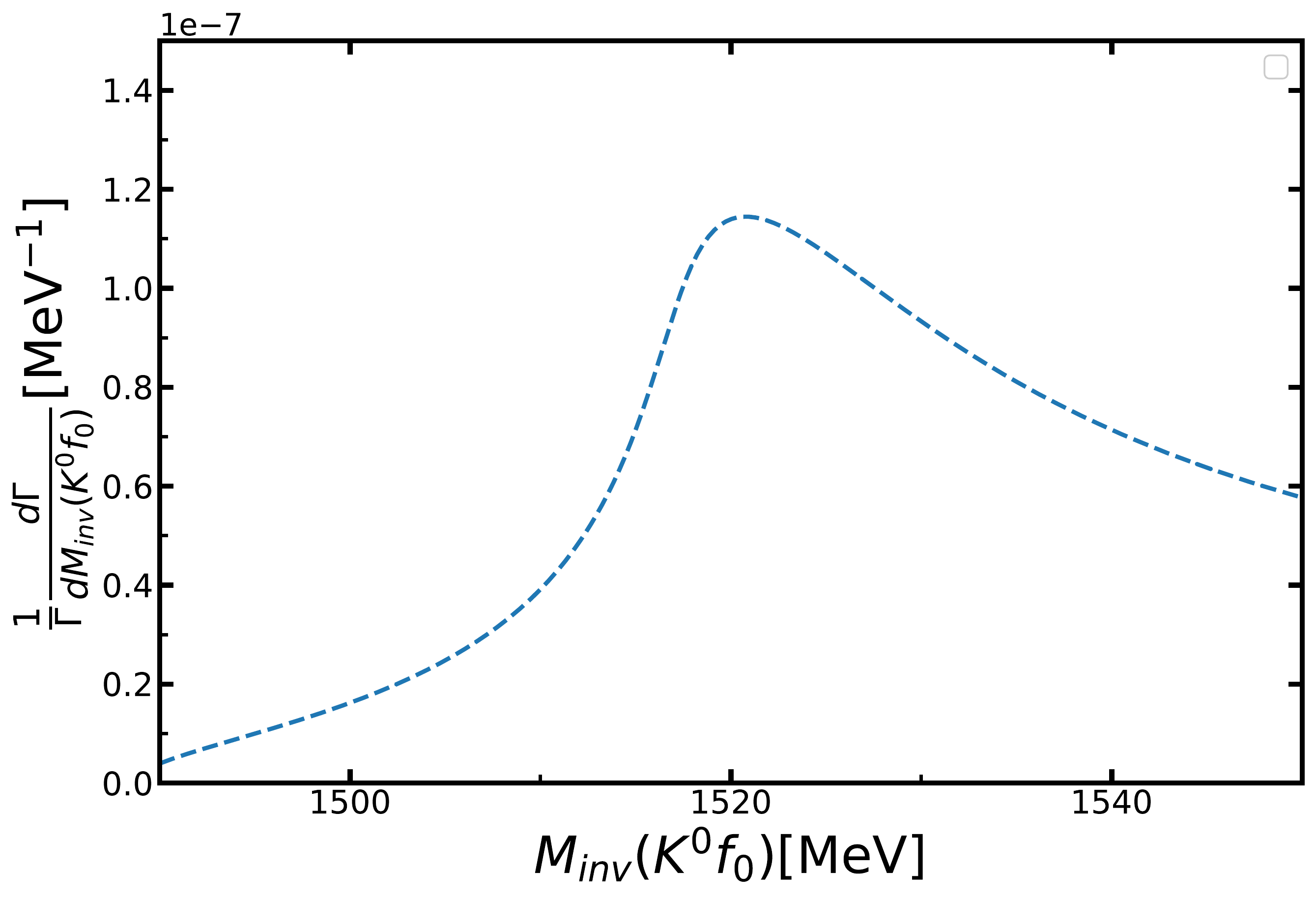}
	\caption{The differential branching ratio $\frac{1}{\Gamma_{B_0}}\frac{d\Gamma_{B_0\to J/\psi K^0 f_0}}{dM_{\rm inv}(K^0f_0)}$ described as in Eq.\,(\ref{1to3}) as a function of $M_{\rm inv}(K^0f_0/K^0a_0)$} 
	\label{fig4} 
\end{figure*}
\begin{figure*}
	\centering 
	\subfigure[]{\includegraphics[width=0.5\textwidth]{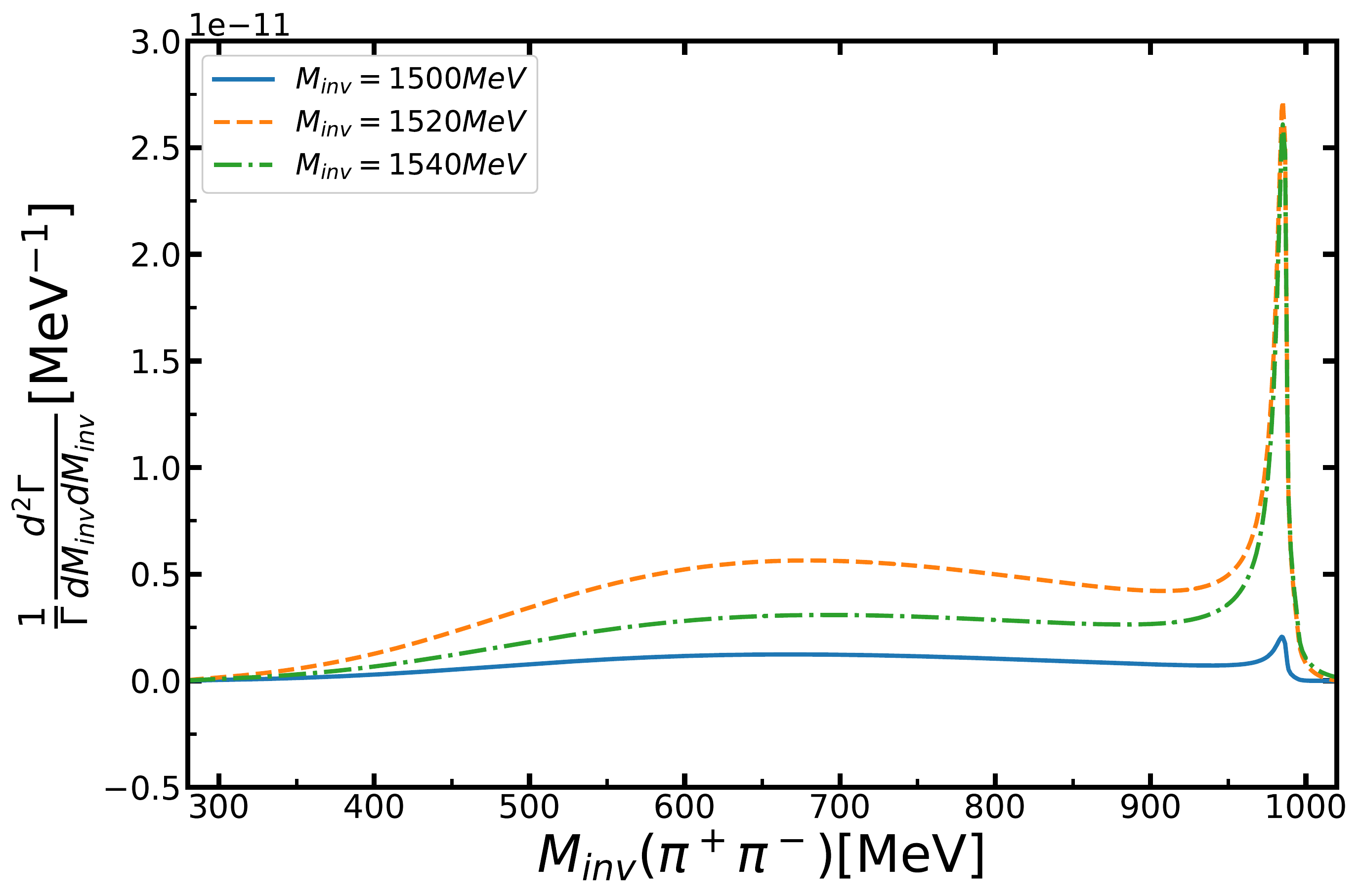}}
	\subfigure[]{\includegraphics[width=0.5\textwidth]{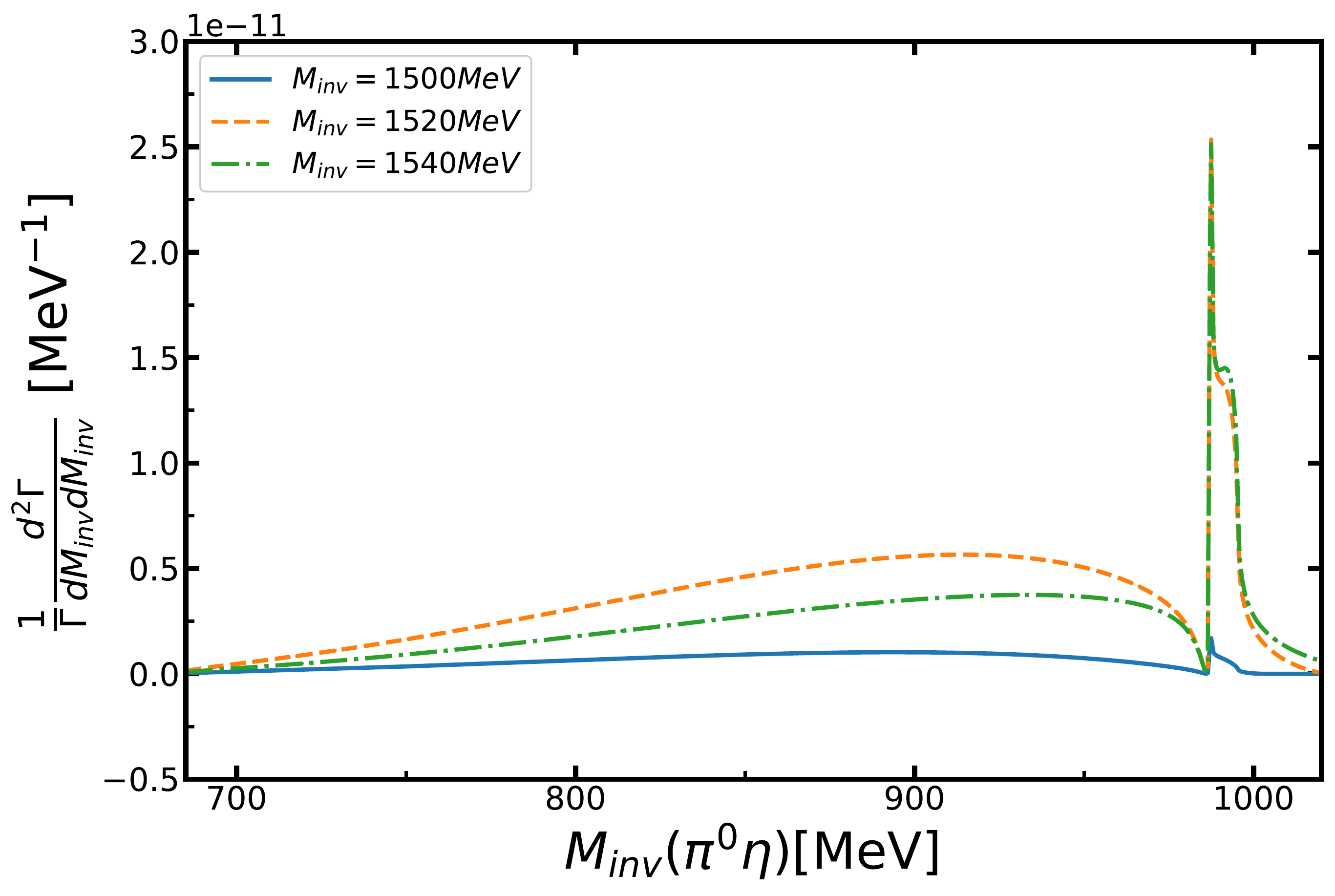}}
	\caption{(a) $\frac{d^2\Gamma_{B_0\to J/\psi K^0f_0(980)\to\pi^+\pi^-}}{dM_{\rm inv}(K^0f_0)dM_{\rm inv}(\pi^+\pi^-)}$ as a function of $M_{\rm inv}(\pi^+\pi^-)$. (b)  $\frac{d^2\Gamma_{B_0\to J/\psi K^0a_0(980)\to J/\psi K^0\pi^0\eta}}{dM_{\rm inv}(K^0a_0)dM_{\rm inv}(\pi^0\eta)}$ as a function of $M_{\rm inv}(\pi^0\eta)$. The blue, orange and green curves are obtained by setting $M_{\rm inv}(K^0f_0)=$1500 MeV, 1520 MeV and 1540 MeV, respectively.}
	\label{fig5}
\end{figure*}
\begin{figure*}
	\centering 
	\subfigure[]{\includegraphics[width=0.4\textwidth]{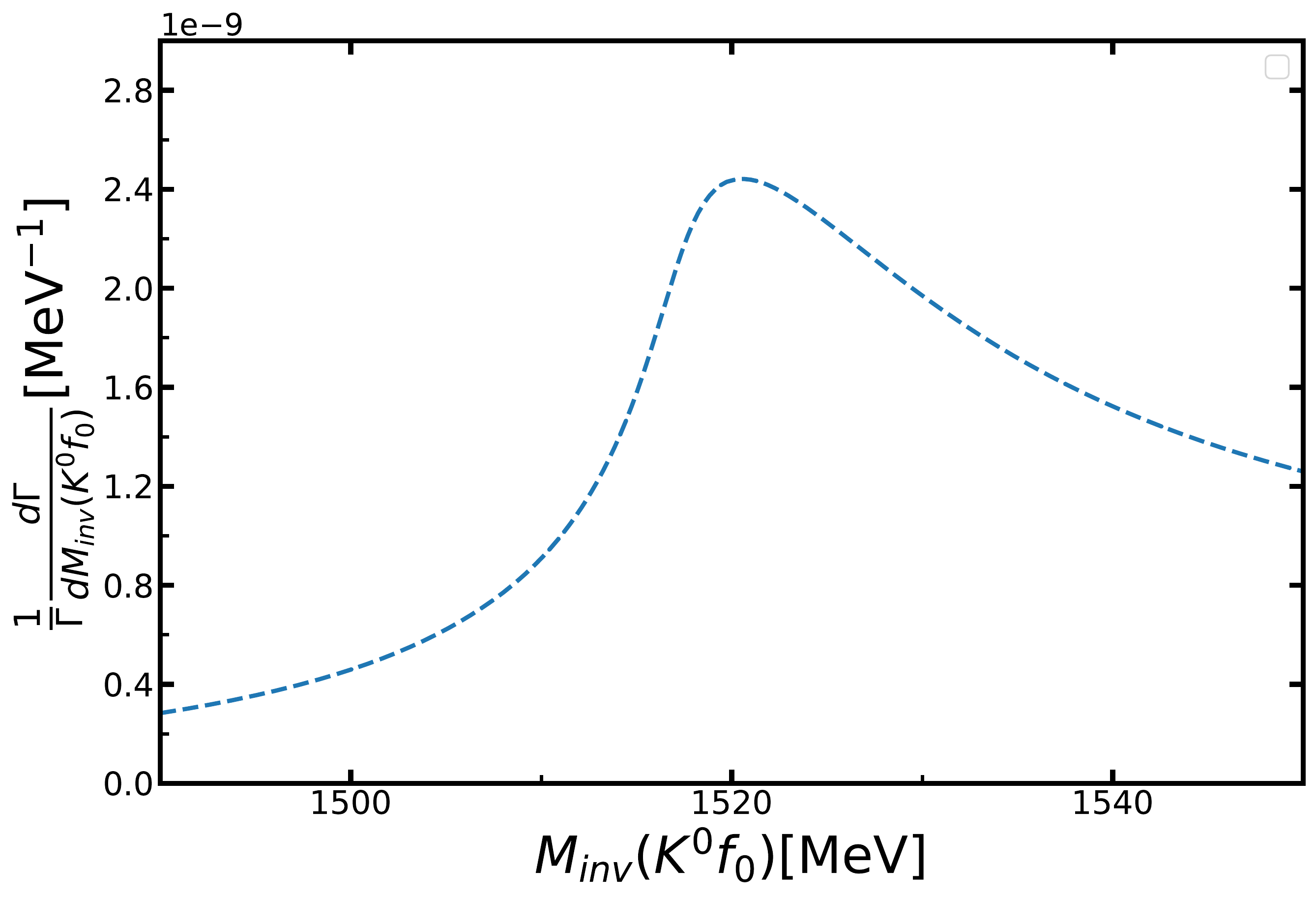}}
	\subfigure[]{\includegraphics[width=0.4\textwidth]{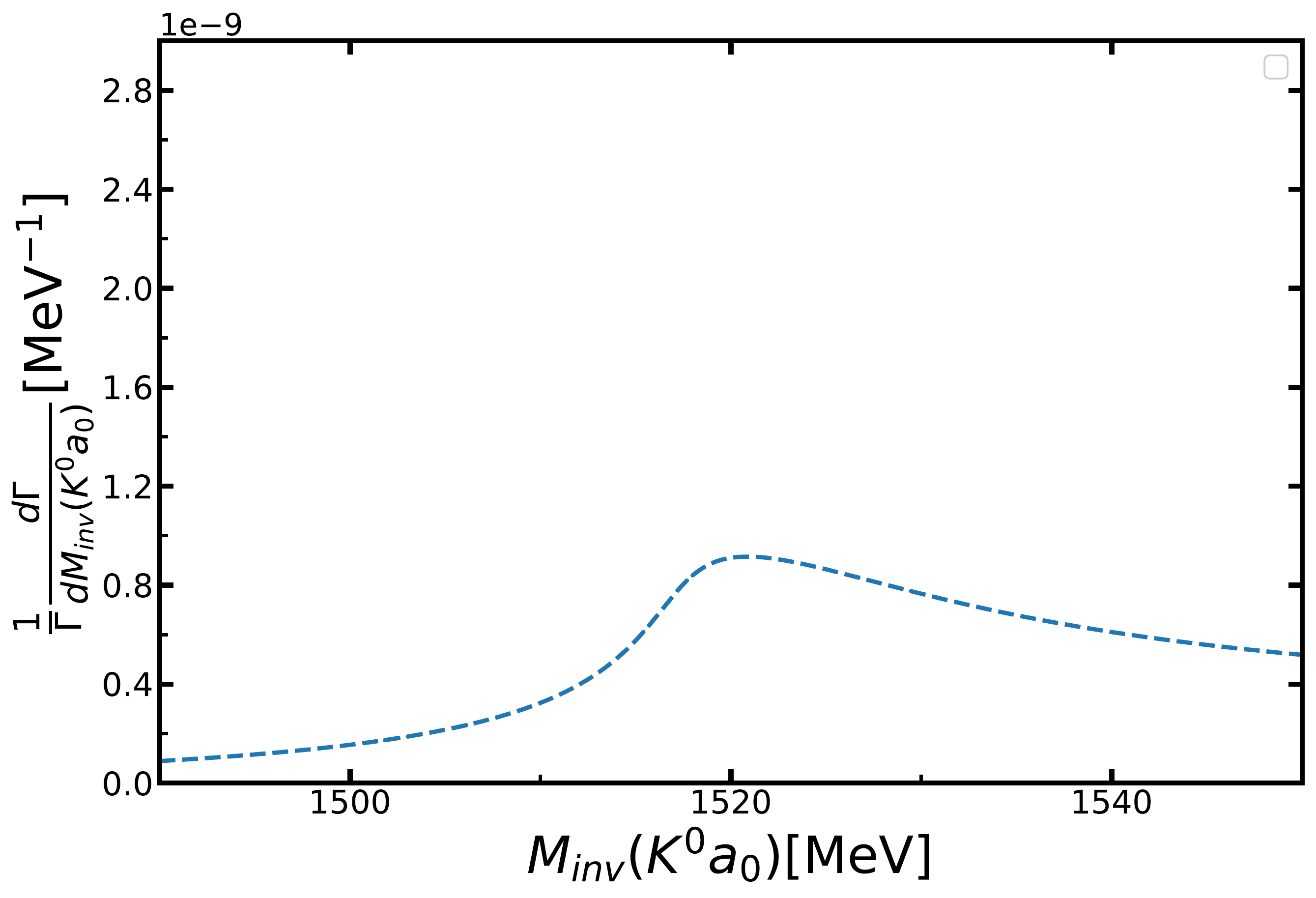}}
	\caption{(a) The differential branching ratio $\frac{d\Gamma_{B_0\to J/\psi K^0f_0(980)\to J/\psi K^0\pi^+\pi^-}}{dM_{\rm inv}(K^0f_0)}$. (b) the branching ratio $\frac{d\Gamma_{B_0\to J/\psi K^0f_0(980)\to J/\psi K^0\pi^+\pi^-}}{dM_{\rm inv}(K^0a_0)}$ described as in Eq.\,(\ref{1to4}) as a function of $M_{\rm inv}(K^0f_0/K^0a_0)$} 
	\label{fig6} 
\end{figure*}
Let us begin by showing in Fig.\,\ref{fig3} the contribution of the triangle loop to the total amplitude and the triangle loop defined in Eq.\,(\ref{tT}). In order to satisfy the TS condition of the Eq.\,(\ref{1}), all intermediate particles must be on the shell so the mass sum of $K^0\bar{K}^0$ must be less than that of $f_0(a_0)$. On the other hand, if the mass of $f_0(a_0)$ is too large, the Eq.\,(\ref{1}) will no longer satisfy. From the perspective of the above considerations, we plot the real and imaginary parts of $t_T$, as well as the absolute value with $M_{\rm inv}(R)$ fixed at 986, 991, and 996 MeV. It can be observed that Re($t_T$) has a peak around 1518 MeV, Im($t_T$) has a peak around 1529 MeV, and there is a peak for $|t_T|$ around 1520 MeV.  As discussed in Refs.\ \cite{Sakai:2017hpg,Dai:2018hqb}, the peak of the imginary part is related to the TSs, while the one of the real part is related to the $K^*K$ threshold. Note that around 1520 MeV and above the TS dominates the reaction.

Then we show $\frac{1}{\Gamma_{J/\psi}}\frac{d\Gamma_{B^0\to J/\psi K^0f_0(a_0)}}{dM_{\rm{inv}}(K^0f_0/K^0a_0)}$. In Fig.\,\ref{fig4} we plot Eq.\,(\ref{1to3}) for  decay process  $B_0\to J/\psi K^0f_0(a_0)$, we also see a peak around 1520 MeV. We can obtain the branching ratio of the 3-body decay process when we integrate over $M_{\rm inv}(K^0R)$,
\begin{align}
	{\rm Br}(B_0\to J/\psi K^0f_0(a_0))=1.007\times10^{-5}.
\end{align}
  
In the upper panel of Fig.\,\ref{fig5} we plot Eq.\,\ref{1to4} for the $B_0\to J/\psi K^0\pi^+\pi^-$ decay, and similarly in the lower panel of Fig.\,\ref{fig5} for the $B_0\to J/\psi K^0\pi^0\eta$ decay as a function of $M_{\rm inv}(R)$, where in both figures we fix $M_{\rm inv}(K^0R)$=1500 MeV, 1520 MeV, and 1540 MeV and vary $M_{\rm inv}(R)$. We can see that the distribution with the largest strength is near $M_{\rm inv}(K^0R)$=1520 MeV, we can also observe a strong peak when $M_{\rm inv}(\pi^+\pi^-)$ around 980 MeV in the upper panel of Fig.\,\ref{fig5}, we can also get very similar results in the lower panel of Fig.\,\ref{fig5}. We see that most of the contribution to the width $\Gamma$ comes from $M_{\rm inv}(K^0R)$=$M_R$, and we have strong contributions for $M_{\rm inv}(\pi^+\pi^-)\in$ [500 MeV, 990 MeV] and $M_{\rm inv}(\pi^0\eta)\in$ [800 MeV, 990 MeV]. Therefore, when we calculate the mass distribution $\frac{d\Gamma}{dM_{\rm inv}(K^0R)}$, we restrict the integral to the limits already mentioned and perform the integration.
\begin{align}
		&\frac{1}{\Gamma_{B_0}} \frac{d \Gamma_{B_0\to J/\psi K^0 f_0(980) \to J/\psi K^0\pi^+\pi^-}}{d M_{\mathrm{inv}}(K^0f_0)} \notag\\
		=&\frac{1}{\Gamma_{J/\psi}} \int_{500 \mathrm{MeV}}^{990 \mathrm{MeV}} dM_{\mathrm{inv}}(\pi^+\pi^-) \notag\\
		&\quad \times \frac{d^{2} \Gamma_{B_0\to J/\psi K^0 f_0(980) \to J/\psi K^0\pi^+\pi^-}}{d M_{\mathrm{inv}}(K^0f_0)d M_{\mathrm{inv}}(\pi^+\pi^-)},
\end{align}
\begin{align}
	&\frac{1}{\Gamma_{B_0}} \frac{d \Gamma_{B_0\to J/\psi K^0 a_0(980) \to J/\psi K^0\pi^0\eta}}{d M_{\mathrm{inv}}(K^0a_0)} \notag\\
	=&\frac{1}{\Gamma_{J/\psi}} \int_{800 \mathrm{MeV}}^{990 \mathrm{MeV}} dM_{\mathrm{inv}}(\pi^0\eta) \notag\\
	&\quad \times \frac{d^{2} \Gamma_{B_0\to J/\psi K^0 a_0(980) \to J/\psi K^0\pi^0\eta}}{d M_{\mathrm{inv}}(K^0a_0)d M_{\mathrm{inv}}(\pi^0\eta)}.
\end{align}

We show the Eq.\,(\ref{1to4}) for both $B_0\to J/\psi K^0\pi^+\pi^-$ and $B_0\to J/\psi K^0\pi^0\eta$. When we integrate over $M_{\rm inv}(R)$ we obtain $\frac{d\Gamma}{dM_{\rm inv}(K^0R)}$ which we show in Fig.\,\ref{fig6}. We see a clear peak of the distribution around 1520 MeV, for $f_0$ and $a_0$ production. At the same time, we can observe that the peak of $M_{\rm inv}(K^0a_0)$ is significantly lower than the peak of $M_{\rm inv}(K^0f_0)$. 

Integrating now $\frac{d\Gamma}{dM_{\rm inv}(K^0f_0)}$ and $\frac{d\Gamma}{dM_{\rm inv}(K^0a_0)}$ over the $M_{\rm inv}(K^0f_0)$($M_{\rm inv}(K^0a_0)$) masses in Fig.\,\ref{fig6}, we obtain the branching fractions
\begin{align}
	&{\rm Br}(B_0\to J/\psi K^0f_0(980)\to J/\psi K^0\pi^+\pi^-)=1.38\times10^{-6},\notag\\
	&{\rm Br}(B_0\to J/\psi K^0a_0(980)\to J/\psi K^0\pi^0\eta)=2.56\times10^{-7}.
\end{align}
	\section{Conclusion}
	\label{IV}
We have performed the calculations for the reactions  $B_0\to J/\psi K^0 f_0(980)(a_0(980))$ and shown that they develop a TS for an invariant mass of 1520 MeV in $(K^0R)$. This TS shows up as a peak in the invariant mass distribution of these pairs with an apparent width of about 20 MeV. We have applied the experimental data of the branching ratio of the decay $B_0\to J/\psi K^0 f_0(980)(a_0(980))$ to determine the coupling strength of the $B_0\to J/\psi K^0 f_0(980)(a_0(980))$ vertex.

We evaluate the $\frac{d^2\Gamma_{total}}{dM_{\rm inv}(K^0R)dM_{\rm inv}(R)}$, and see clear peaks in the distributions $M_{\rm inv}(\pi^+\pi^-)$($M_{\rm inv}(\pi^0\eta)$), showing clearly the $f_0(a_0)$ shapes. Integrating over $M_{\rm inv}(R)$ respectively,   these distributions show a clear peak for $M_{\rm inv}(K^0R)$ around 1520 MeV. 

This peak is a result of the singularity of the triangle and may be misidentified with resonance when the experiment is completed. In this sense, the work done here should serve as a warning not to treat it as resonance when this peak is seen in future experiments. It is important to discover new conditions about TSs, and to allow for this possibility when experimentally observed peaks can avoid associating these peaks with resonance. The value of this work lies in identifying a TS for a suitable reaction and then preparing the results and research to interpret the peak correctly when it is observed.
	\section{Acknowledgments}
	\label{V}
This work is partly supported by the National Natural Science Foundation of China under Grants No. 12205002, and partly supported by the the Natural
Science Foundation of Anhui Province (No.2108085MA20). 

\bibliography{submit}	
\end{document}